\documentclass[aps,pra,twocolumn,showpacs,superscriptaddress,longbibliography]{revtex4-2}
\usepackage{graphicx} 
\usepackage{amsmath}
\usepackage{graphicx,epstopdf}
\usepackage[flushleft]{threeparttable}
\usepackage{gensymb}
\newcommand{\be}{\begin{equation}}
\newcommand{\ee}{\end{equation}}
\newcommand{\bea}{\begin{eqnarray}}
\newcommand{\eea}{\end{eqnarray}}
\newcommand{\bse}{\begin{subequations}}
\newcommand{\ese}{\end{subequations}}

\usepackage{color}
\usepackage[colorlinks,bookmarks=false,citecolor=darkblue,linkcolor=red,urlcolor=blue]{hyperref}

\definecolor{darkred}{rgb}{0.7,0.0,0.0}

\definecolor{darkblue}{rgb}{0,0.02,0.45}

\definecolor{darkgreen}{rgb}{0.02,0.45,0.0}

\definecolor{violet}{rgb}{0.8,0.2,0.6}

\begin{document}

\title{Magnetic properties of frustrated spin-$\frac{1}{2}$ capped-kagome antiferromagnet (CsBr)Cu$_5$V$_2$O$_{10}$}
\author{S. Guchhait}
\affiliation{School of Physics, Indian Institute of Science Education and Research, Thiruvananthapuram 695551, India}
\author{D. V. Ambika}
\affiliation{Ames National Laboratory, U.S. DOE, Ames, Iowa 50011, USA}
\affiliation{Department of Physics and Astronomy, Iowa State University, Ames, Iowa 50011, USA}
\author{S. Mohanty}
\affiliation{School of Physics, Indian Institute of Science Education and Research, Thiruvananthapuram 695551, India}
\author{Y. Furukawa}
\affiliation{Ames National Laboratory, U.S. DOE, Ames, Iowa 50011, USA}
\affiliation{Department of Physics and Astronomy, Iowa State University, Ames, Iowa 50011, USA}
\author{R. Nath}
\email{rnath@iisertvm.ac.in}
\affiliation{School of Physics, Indian Institute of Science Education and Research, Thiruvananthapuram 695551, India}

\date{\today}

\begin{abstract}
The structural and magnetic properties of a spin-$\frac{1}{2}$ averievite (CsBr)Cu$_5$V$_2$O$_{10}$ are investigated by means of temperature-dependent x-ray diffraction, magnetization, heat capacity, and $^{51}$V nuclear magnetic resonance (NMR) measurements. The crystal structure (trigonal, $P\bar{3}$) features a frustrated capped-kagome lattice of the magnetic Cu$^{2+}$ ions. Magnetic susceptibility analysis indicates a large Curie-Weiss temperature of $\theta_{\rm CW} \simeq-175$~K. Heat capacity signals the onset of a magnetic long-range-order (LRO) at $T_{\rm N}\simeq 21.5$~K at zero magnetic field due to the presence of significant inter-planer coupling in this system. The magnetic LRO below 27~K is further evident from the drastic change in the $^{51}$V NMR signal intensity and rapid enhancement in the $^{51}$V spin-lattice relaxation rate in a magnetic field of 6.3~T. The frustration index $f=|\theta_{\rm CW}|/T_{\rm N} \simeq 8$ ascertains strong magnetic frustration in this compound. From the high-temperature value of the $^{51}$V NMR spin-lattice relaxation rate, the leading antiferromagnetic exchange interaction between the Cu$^{2+}$ ions is calculated to be $J/k_{\rm B}\simeq 136$~K. 
\end{abstract}

\maketitle

\section{Introduction}
Low-dimensional and frustrated magnets furnish a fertile ground to enrich the knowledge on phase transitions in quantum magnetism. In such materials, ground state degeneracy along with strong quantum fluctuations melt the magnetic long-range-order (LRO), leading to highly entangled and quantum disordered ground states, such as quantum spin-liquid (QSL)~\cite{Balents199,Ramirez453,Zhou025003,Savary2017}. Triangular lattice antiferromagnet is one of the most familiar examples of two-dimensional (2D) geometrically frustrated magnets, where the magnetic ions are embedded at the vertices of the edge shared triangular motifs. The kagome Heisenberg antiferromagnet (KHAF) is another type of geometrically frustrated 2D lattice in which the triangular motifs are corner-shared and the magnetic frustration is magnified compared to the edge-shared triangular lattice. Further, the effect of quantum fluctuations in a KHAF is expected to be enhanced for low spin systems (e.g. $S=1/2$) and such systems are well suited for studying quantum phase transitions~\cite{Yan1173}. A well-studied example is herbertsmithite ZnCu$_3$(OH)$_6$Cl$_2$, which exhibits neither spin-freezing nor magnetic ordering down to 50~mK, confirming QSL behavior~\cite{Helton107204,Han406}.

Depending on the sign and relative strength of the exchange couplings, several phase diagrams with different ground states are reported theoretically for KHAFs~\cite{Suttner020408,Boldrin220408,Colbois174403,Li035112}. Similarly, depending on the value of spin quantum number, multitude of magnetization plateaus corresponding to field-induced spin nematic, simplex solid, supersolid, superfluid, Valance Bond Crystal, and QSL state in isothermal magnetization curve are also delineated~\cite{Nishimoto1,Picot060407,FangL220401}.
A striking feature of KHAF is that one can manipulate the lattice geometry and design many derivatives of kagome lattice such as octa-kagome~\cite{Peng075140,Tang14057}, staircase-kagome~\cite{Yoon214401,Morosan144403,Liu170228}, sphere-kagome~\cite{Rousochatzakis094420}, strip-kagome~\cite{Jeschke140410, Morita161}, hyper-kagome~\cite{Zhao014441, SimutisL100404, Okamoto137207}, tripod-kagome~\cite{Dun157201, Dun031069, Dun104439}, square-kagome~\cite{Fujihala3429, Liu155153}, and breathing-kagome~\cite{Flavian174406}. Because of complex lattice geometry, these compounds have the potential to host numerous nontrivial ground states. To name a few, QSL is realized in the square-kagome compound KCu$_6$AlBiO$_4$(SO$_4$)$_5$Cl~\cite{Fujihala3429,Liu155153} and hyperkagome compound Na$_4$Ir$_3$O$_8$~\cite{Okamoto137207} and spin-ice-type ground state is reported in the tripod-kagome compound Mg$_2$Dy$_3$Sb$_3$O$_{14}$~\cite{Dun157201}. 

The Cu$^{2+}$ based averievite family with general formula ($MX$)Cu$_5$O$_2$($T^{5+}$O$_4$)$_2$ [$M$ = K, Rb, Cs, Cu; $X$ = Cl, Br; and $T^{5+}$ = P, V] are a new class of compounds with a highly frustrated capped-kagome geometry~\cite{Kornyakov1833}. Recently, the magnetic properties of two compounds (CsCl)Cu$_5$V$_2$O$_{10}$ (CCCVO) and (RbCl)Cu$_5$P$_2$O$_{10}$ (RCCPO) of this family are reported in detail~\cite{Botana054421,Mohanty104424}. Majority of the compounds in this series undergo a structural transition from high symmetric trigonal at high temperatures to low symmetric monoclinic at low temperatures. For both CCCVO and RCCPO, the structural transition is reported at around $T_{\rm t} \simeq 310$~K. The crystal structure of all these compounds comprises OCu$_4$ tetrahedra, where Cu$^{2+}$ ions form a 2D capped-kagome layer. Despite very large Curie-Weiss (CW) temperature ($\theta_{\rm CW} \simeq-185$~K for CCCVO and -160~K for RCCPO), both the compounds undergo magnetic ordering ($T_{\rm N}$) at very low temperatures. Unlike CCCVO, which has one antiferromagnetic (AFM) transition at $T_{\rm N}\simeq 24$~K, RCCPO undergoes two magnetic transitions at low temperatures, possibly driven by anisotropy~\cite{Mohanty104424}. Moreover, CCCVO exhibits another structural transition at low temperatures, which may be due to the ordering of Cs atoms and such a transition is absent for RCCPO. The Zn$^{2+}$ substitution in place of capped Cu$^{2+}$ ion in CCCVO suppresses the magnetic LRO as well as the structural transition, making it a promising candidate for QSL~\cite{Botana054421,BiesnerL060410,Georgopoulou2023}.
Some preliminary magnetic measurements of the phosphate analog compounds of CCCVO [i.e. Cs(Cl,Br,I)Cu$_5$P$_2$O$_{10}$] are also reported~\cite{Winiarski58,Dey125106}.
Moreover, the recent theoretical studies have proposed that the spin-$1/2$ capped-kagome systems may show different magnetization plateaus at $1/5$, $3/5$, and $4/5$ of the saturation magnetization due to the formation of localized magnons, singlets, and doublets, respectively~\cite{Rausch143,Zhang2299}.

\begin{figure}
\includegraphics[scale=0.6]{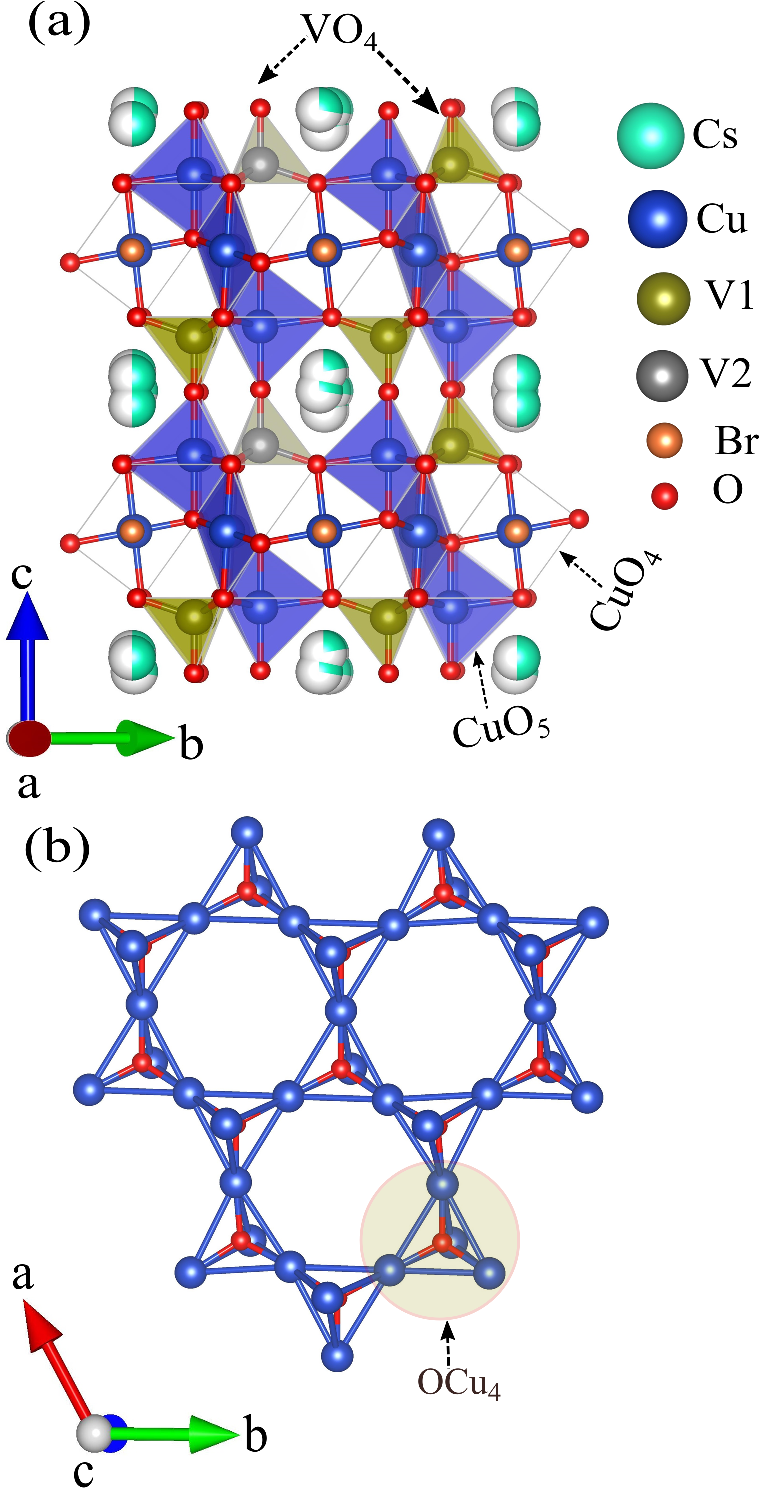} 
\caption{(a) Crystal structure of (CsBr)Cu$_5$V$_2$O$_{10}$ projected in the $bc$-plane. The connection of Cu(3,4)O$_4$ square planes and Cu(1,2)O$_5$ trigonal bipyramids form the capped-kagome layers. Adjacent layers are connected through VO$_4$ tetrahedra along the $c$-direction. (b) A section of capped-kagome layer of Cu$^{2+}$ ions in the $ab$-plane.}
    \label{Fig1}
\end{figure}

Herein, we report the synthesis and a detailed study of the structural and magnetic properties of another capped-kagome compound (CsBr)Cu$_5$V$_2$O$_{10}$ (abbreviated as CBCVO), belonging to the same averievite family. It crystallizes in the trigonal symmetry with space group$P\bar{3}$ at room temperature, without any structural transition. The crystal structure of CBCVO is presented in Fig.~\ref{Fig1}. There are four inequivalent Cu sites and two inequivalent V sites in the crystal structure. Among them, Cu(3) and Cu(4) form distorted CuO$_4$ square planner units while Cu(1) and Cu(2) form distorted CuO$_5$ trigonal bipyramids. The corner shared CuO$_4$ units constitute a kagome layer in the $ab$-plane. The CuO$_5$ units share the edges with the neighbouring CuO$_4$ units, building the distorted OCu$_4$ tetrahedra. The corner-shared OCu$_4$ oxocentered plaquettes are arranged in an up-down-up-down manner and form a 2D capped-kagome layer in the $ab$-plane [see Fig.~\ref{Fig1}(b)]. Disordered Cs$^{+}$ ions are positioned between the adjacent capped-kagome layers while the Br$^{-}$ ions are located in the middle of the hexagonal rings. Similar to the other members of the averievite family, CBCVO has a large value of $\theta_{\rm CW} \simeq -175$~K and it shows the onset of a magnetic LRO at $T_{\rm N} \simeq 21.5$~K at zero magnetic field. With a frustration index of $f \simeq 8$, this compound is categorized to be a strongly frustrated magnet.



\section{Methods}
Polycrystalline sample of CBCVO was synthesized by the conventional solid-state reaction technique. High pure starting materials CsBr (Sigma-Aldrich, 99.99\%), CuO (Sigma-Aldrich, 99.99\%), and V$_2$O$_5$ (Sigma-Aldrich, 99.99\%) in powder form were mixed in a stoichiometric proportion. The mixture was ground well into fine powder with the help of an agate mortar and pelletized. The pellets were sealed in an evacuated (10$^{-2}$~mbar) quartz tube and sintered at 400$\degree$C for 12~h followed by slow cooling to 350$\degree$C in a cooling rate of 0.5$\degree$C/min. Finally, the furnace was switched off and the sample was furnace cooled to room temperature. To confirm the phase purity and to detect the structural transition, powder x-ray diffraction (PXRD) measurement was performed using a PANalytical (Cu$K_\alpha$ radiation, $\lambda_{\rm avg}\simeq1.54060$~\AA) diffractometer as a function of temperature (13~K$\leq T\leq 600$~K). For low- and high-temperature measurements, a low-temperature attachment (Oxford PheniX) and a high-temperature oven attachment (Anton-Paar HTK 1200N) to the x-ray diffractometer were used.

Magnetization ($M$) measurements were performed using the vibrating sample magnetometer (VSM) attachment to the Physical Property Measurement System (PPMS, Quantum Design) as a function of temperature (2~K$\leq T\leq 600$~K) and magnetic field ($0 \leq H \leq 9$~T). For high-temperature ($T \geq 380$~K) measurements, a high-$T$ oven (Model: CM-C-VSM) attachment to the VSM was used. Temperature-dependent heat capacity [$C_{\rm p}(T)$] in zero field was measured down to 2~K in PPMS by the thermal relaxation technique. 

The nuclear magnetic resonance (NMR) measurements were performed on the powder sample and on the $^{51}$V nucleus which has nuclear spin $I=\frac{7}{2}$ and gyromagnetic ratio $^{51}\gamma_{\rm N}/2\pi = 11.193$~MHz/T. We carried out these experiments using a homemade phase-coherent spin-echo pulse spectrometer in the temperature range 4.3~K$\leq T\leq 250$~K. We used a silver coil instead of a copper coil to avoid any Cu signals from the NMR coil. NMR spectra at different temperatures were measured by varying the external magnetic field at a fixed radio frequency $\nu = 71.8$~MHz ($\nu = \gamma_{\rm N} H$). The nuclear spin-lattice relaxation rate ($1/T_1$) was measured using the conventional single saturation pulse method at the central peak position of the NMR spectrum. Temperature variation of NMR shift, $K (\%) = \left[\frac{H_{\rm 0}}{H}-1\right]$, was calculated from the resonance field of the sample ($H$) with respect to the Larmor field (zero shift position: $H_{\rm 0}$).

\section{Results and Discussion}
\subsection{X-ray Diffraction}
\begin{figure}
\includegraphics[width=\linewidth]{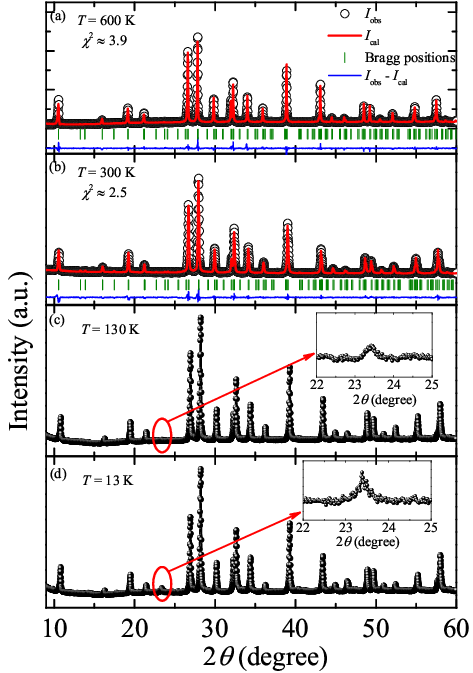}
\caption{Powder XRD pattern (open circles) at (a) 600~K and (b) 300~K along with the Rietveld refinement (solid line). The Bragg peak positions are indicated by green vertical bars while the lower solid line corresponds to the difference between the experimental and calculated intensities. (c) and (d) represent the powder XRD patterns at 130 and 13~K, respectively. Insets highlight the extra peak at around $2\theta\simeq 23.4^{\degree}$ that appears below 130~K.}
\label{Fig2}
\end{figure}
\begin{figure}
\includegraphics[width=\linewidth]{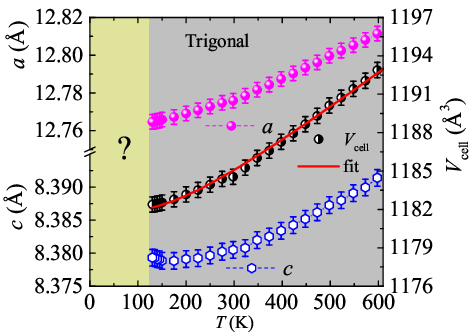}
\caption{Temperature dependence of lattice parameters ($a$ and $c$) and unit cell volume ($V_{\rm cell}$) from 600~K to 130~K. The solid line represents the fit of $V_{\rm cell}(T)$ by Eq.~\eqref{Vcell}.}
\label{Fig3}
\end{figure}
Figure~\ref{Fig2}(b) presents the powder XRD pattern of CBCVO at room temperature. Rietveld refinement of the XRD pattern was performed using the FULLPROF program~\cite{Carvajal55}, taking the initial structural parameters of CCCVO~\cite{Kornyakov1833}. All the peaks could be modeled via Rietveld refinement using the trigonal crystal structure with $P\bar{3}$ symmetry. This confirms CBCVO is isostructural to CCCVO~\cite{Kornyakov1833}. However, our refinement using trigonal crystal structure with space group $P\bar{3}m1$ as reported in Ref.~\cite{Botana054421} fails to model the XRD data adequately. The obtained atomic parameters are listed in Table~\ref{Refined parameters}. The refined lattice constants and unit cell volume ($V_{\rm cell}$) are $a=12.773(5)$~\AA, $c=8.380(3)$~\AA, and $V_{\rm cell} \simeq 1183.6$~\AA$^3$, respectively which are slightly larger than the values of CCCVO [$a=12.724(4)$~\AA, $c=8.376(4)$~\AA, and $V_{\rm cell}\simeq 1174.6$~\AA$^3$]. This may be due to the larger ionic radius of Br$^{-}$ than that of Cl$^{-}$. The refined structure of CBCVO is presented in Fig.~\ref{Fig1}.

Powder XRD was also measured in different intermediate temperatures between 13~K and 600~K. All the XRD patterns down to 130~K are found to be identical to the room temperature XRD pattern, indicating no structural distortion. Surprisingly, unlike CCCVO, no structural transition from trigonal to monoclinic is detected for CBCVO. However, below 130~K an extra peak appears at around $2\theta \simeq 23.4\degree$ and that persists down to 13~K with increased intensity [see Fig~\ref{Fig2}(c) and (d)]. Though the origin of this peak is not yet understood, same type of a unidentified peak is also reported for CCCVO below 127~K, which is ascribed to the ordering of the Cs sites~\cite{Botana054421}.

The temperature dependence of the lattice parameters ($a$ and $c$) and $V_{\rm cell}$ in the trigonal phase are presented in Fig.~\ref{Fig3}. The lattice constants $a$ and $c$ are found to decrease in a systematic way with decreasing temperature, which leads to an overall thermal contraction of $V_{\rm cell}$ with temperature. The temperature evolution of $V_{\rm cell}$ can be expressed in terms of the internal energy [$U(T)$] of the system~\cite{Guchhait224415},
\begin{equation}\label{Vcell}
V_{\rm cell}(T) = \frac{\gamma U(T)}{K_0} + V_0.
\end{equation}
Here, $V_0$ is the unit-cell volume at $T = 0$~K, $\gamma$ is the Gr\"uneisen parameter, and $K_0$ is the bulk modulus of the system. According to the Debye model, $U(T)$ can be written as,
\begin{equation}\label{Uenergy}
U(T) = 9Nk_{\rm B}T\left(\frac{T}{\theta_{\rm D}}\right)^3 \int_{0}^{\frac{\theta_{\rm D}}{T}}\frac{x^3}{(e^{x}-1)}dx,
\end{equation}
where $N$ is the total number of atoms per unit cell, $k_{\rm B}$ is the Boltzmann constant, and $\theta_{\rm D}$ is the Debye temperature~\cite{Kittel2004}. The variable $x$ stands for the quantity $\frac{\hbar\omega}{k_{\rm B}T}$, where $\omega$ is the phonon frequency and $\hbar$ is the reduced Planck constant. Here, $\theta_{\rm D}=\frac{\hbar\omega_{\rm D}}{k_{\rm B}}$ and $\omega_{\rm D}$ is the cutoff frequency. The best fit of the $V_{\rm cell}(T)$ data down to 130~K using Eq.~\eqref{Vcell} [solid line in Fig.~\ref{Fig3}] results in the parameters: $\theta_{\rm D} = 700(8)$~K, $V_0 = 1181.77(5)$~\AA$^3$, and $\frac{\gamma}{K_0} \simeq 9.194\times10^{-12}$~Pa$^{-1}$. 
 
\begin{table}[ptb]
	\caption{Structural parameters of CBCVO obtained from the Rietveld refinement of the powder-XRD data at 300~K [Trigonal, Space group: $P\bar{3}$ (No. 147)]. Our fit yields $a=12.773(5)$~\AA, $c=8.380(3)$~\AA, and $V_{\rm cell} \simeq 1183.6$~\AA$^3$. Listed are the Wyckoff positions and the refined atomic coordinates for each atom.}
	\label{Refined parameters}
        \begin{ruledtabular}
        \begin{tabular}{ccccccc}	
	Atoms & Wyckoff & $x$ & $y$ & $z$ & Occ.
 \\\hline
		Cs1 & 2$c$ &0.498(1) & 0.519(4)& 0.000 & 0.28 \\
		Cs2 & 6$g$ & 0.553(2) & 0.517(5) & 0.047(1) & 0.22 \\
		Cs3 & 6$g$ & 1.000 & 1.000 & 0.043(3) & 0.5\\
		Cu1 & 2$d$ & 0.333 & 0.666& -0.223(1) & 1.00\\
	    Cu2 & 6$g$ & 0.666 & 0.828(2) & 0.229(6) & 1.00\\
	    Cu3 & 6$g$ & 0.741(5) & 0.992(7) & 0.484(1) & 1.00\\
	    Cu4 & 6$g$ & 0.487(3) & 0.746(4) & 0.504(2) & 1.00\\
	    V1 & 6$g$ & 0.682(5) & 0.846(1) & -0.189(3) & 1.00\\
	    V2 & 2$d$ & 0.333 & 0.666 & 0.194(7) & 1.00\\
	    Br1 & 1$b$ & 1.000 & 1.000 & 0.500 & 1.00\\
	    Br2 & 3$f$ & 0.500 & 0.500 & 0.500& 1.00\\
	    O1 & 6$g$ & 0.749(4) & 0.980(6) & -0.250(2) & 1.00\\
	    O2 & 6$g$ & 0.753(5) & 1.020 (9) & 0.234 (3) & 1.00\\
	    O3 & 6$g$ & 0.508(3) & 0.748(4) & -0.269(5) & 1.00\\
	    O4 & 6$g$ & 0.641(3) & 0.794(2) & 0.027(2) & 1.00\\
	    O5 & 2$d$ & 0.333 & 0.666 & 0.035(1) & 1.00\\
	    O6 & 6$g$ & 0.485(8) & 0.747(7) & 0.265(3) & 1.00\\
	    O7& 2$d$ & 0.333 & 0.666 & 0.491(6) & 1.00\\
	    O8 & 6$g$ & 0.675(2) & 0.818(2) & 0.451(1) & 1.00\\
   \end{tabular}
   \end{ruledtabular}
\end{table} 

\subsection{Magnetization}
\begin{figure}
	\includegraphics[width=\linewidth]{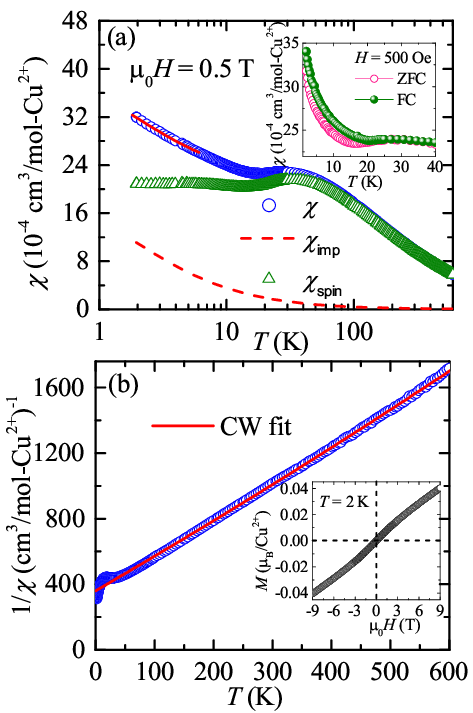}
	\caption{(a) $\chi$ vs $T$ in an applied field of $\mu_{\rm 0}H = 0.5$~T. The solid line is the Curie fit ($\chi_0 + C_{\rm imp}/T$) in the low temperature regime, as described in the text. The dashed line represents the Curie ($\chi_{\rm imp} = C_{\rm imp}/T$) contribution. $\chi_{\rm spin}(T)$ obtained after substracting $C_{\rm imp}/T$ from $\chi(T)$ is also shown. Inset: $\chi(T)$ measured at $\mu_{\rm 0}H = 500$~Oe in ZFC and FC protocols. (b) $1/\chi$ vs $T$ at $\mu_{\rm 0}H = 0.5$~T and the solid line is the CW fit. Inset: $M$ vs $H$ from -9 to 9~T at $T = 2$~K.}
	\label{Fig4}
\end{figure}
Temperature dependent magnetic susceptibility $\chi$ ($\equiv M/H$) of CBCVO measured in a magnetic field of $\mu_{0}H = 0.5$~T is shown in Fig.~\ref{Fig4}(a). In the high-temperature regime, $\chi(T)$ follows a Curie-Weiss (CW) behavior and passes through a broad maxima at $T_{\chi}^{\rm max}\simeq 35$~K. This broad maxima is a signature of the evolution of short-range magnetic correlations among the Cu$^{2+}$ ions, as expected for a 2D antiferromagnet. For $T < 10$~K, $\chi(T)$ shows a low-$T$ upturn, typical for powder samples due to the presence of orphan spins or lattice defects. No sign of any magnetic LRO was observed down to 2~K. However, $\chi(T)$ measured in zero-field cooled (ZFC) and field-cooled (FC) protocols in a small magnetic field $\mu_0 H = 500$~Oe [inset of Fig.~\ref{Fig4}(a)] displays a weak bifurcation at around $T_{\rm N} \simeq 21$~K. This could be a possible indication of either the onset of a magnetic LRO or freezing of a small fraction of spins due to site disorder.

In Fig.~\ref{Fig4}(b), we have plotted the inverse susceptibility ($1/\chi$) as a function of temperature for $\mu_{0}H = 0.5$~T. It exhibits a perfectly linear behaviour at high-temperatures and can be fitted by the modified CW law
\begin{equation}\label{CW}
\chi(T) = \chi_0 + \frac{C}{T - \theta_{\rm CW}}.
\end{equation}
Here, $\chi_0$ is the temperature-independent susceptibility consisting of core diamagnetic susceptibility ($\chi_{\rm dia}$) of the core electron shells of the atoms and Van-Vleck paramagnetic susceptibility ($\chi_{\rm VV}$) of the open shells of the Cu$^{2+}$ ions in the sample. In the second term, $C$ and $\theta_{\rm CW}$ are the Curie constant and CW temperature, respectively. The fit for $T \geq 370$~K yields $\chi_0 \simeq -5.18 \times 10^{-5}$~cm$^3$/mol-Cu$^{2+}$, $C \simeq 0.49$~cm$^3$.K/mol-Cu$^{2+}$, and $\theta_{\rm CW} \simeq -175$~K. Such a large and negative value of $\theta_{\rm CW}$ indicates that the dominant exchange coupling between the Cu$^{2+}$ ions is AFM in nature. This $C$ value corresponds to an effective magnetic moment $\mu_{\rm eff}\simeq 1.99$~$\mu_{\rm B}/$Cu$^{2+}$, implying a $g$-factor of $\sim 2.29$. A slightly larger value of $g$ ($ > 2 $) is typically found for Cu$^{2+}$ based oxides due to a minuscule spin-orbit interaction~\cite{Nath014407,Jin174423}.
The value of $\chi_{\rm dia}$ was estimated to be $-2.5 \times 10^{-4}$~cm$^3$/mol by summing the core diamagnetic susceptibility of each ion (Cs$^{+}$, Br$^{-}$, Cu$^{2+}$, V$^{5+}$, and O$^{2-}$)~\cite{Selwood2013}. $\chi_{\rm VV}$ was calculated by subtracting $\chi_{\rm dia}$ from $\chi_0$ to be $\sim 1.98 \times10^{-4}$~cm$^3$/mol, which is close to the value reported for other Cu$^{2+}$ based compounds~\cite{Guchhait224415,Islam174432}.

Further, to make a tentative estimation of the extrinsic paramagnetic contributions, the $\chi(T)$ data below 6~K in Fig.~\ref{Fig4}(a) was fitted using $\chi = \chi_0 + C_{\rm imp}/T$. Here, the second term is the Curie law. The obtained Curie constant $C_{\rm imp}\simeq 0.00417$~cm$^3$ K/mol corresponds to an impurity concentration of nearly $\sim 1.$\%, assuming impurity spins $S = 1/2$. The intrinsic spin susceptibility $\chi_{\rm spin}$ was estimated by substracting $C_{\rm imp}/T$ from $\chi$ and plotted as a function of temperature in the same graph.

From the experimental value of $\theta_{\rm CW}$, one can estimate the average value of the AFM exchange coupling as $\theta_{\rm CW}=[-zJS(S+1)]/3k_{\rm B}$~\cite{Domb296,Lal014429}. Here, $J$ is the nearest-neighbour(NN) exchange coupling with the Heisenberg Hamiltonian $H = J \sum S_i \cdot S_j$ and $z$ is the number of NN spins of each Cu$^{2+}$ ion. In capped-kagome layer, each Cu$^{2+}$ ion in the kagome plane is connected with six NN ions while each capped Cu$^{2+}$ ion has only three NN ions. Hence, on average each Cu$^{2+}$ spin interacts with 4.5 neighbouring spins. Thus, using the value of $\theta_{\rm CW}$, $z=4.5$, and $S=1/2$ in the above expression, we estimated $J/k_{\rm B} \simeq 155$~K. An interesting feature of this compound is that the broad maximum occurs at a much reduced temperature, $T_{\chi}^{\rm max} \simeq 0.22 J/k_{\rm B}$, compared to $\theta_{\rm CW}$. This is a preliminary signature of strong magnetic frustration in the compound~\cite{Quilliam180401}.

A magnetic isotherm ($M$ vs $H$) measured at $T=2$~K is shown in the inset of Fig.~\ref{Fig4}(b). $M$ changes linearly with $H$, a typical behavior of an antiferromagnet and the absence of a hysteresis around zero-field rules out a spin-glass type phase. However, it doesn't preclude the freezing of a small fraction of spins.

\subsection{Heat Capacity}
\begin{figure}
	\includegraphics[width=\linewidth]{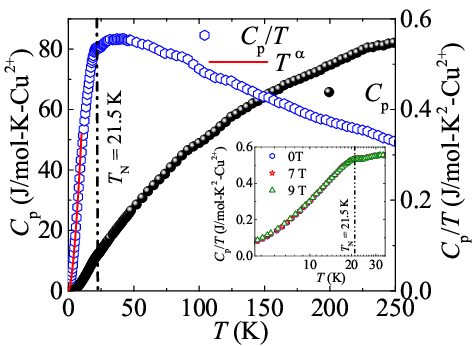}
	\caption {$C_{\rm P}$ (left $y$-axis) and $C_{\rm p}/T$ (right $y$-axis) vs $T$ measured in zero magnetic field. The vertical dashed line indicates the magnetic LRO at $T_{\rm N} \simeq 21.5$~K. The red solid line is the power law fit to the $C_{\rm p}$ data well below $T_{\rm N}$. Inset: Magnified $C_{\rm p}/T$ vs $T$ data around $T_{\rm N}$ measured in different fields.}
	\label{Fig5}
\end{figure}
The zero-field heat capacity ($C_{\rm p}$) as a function of temperature measured down to 1.9~K is shown in Fig.~\ref{Fig5}. The right $y$-axis of Fig.~\ref{Fig5} shows the temperature dependence of $C_{\rm P}/T$. A small kink is observed at around $T_{\rm N} \simeq 21.5$~K that coincides with the bifurcation temperature of the ZFC and FC $\chi(T)$ curves. This implies a transition to the magnetic LRO state. The $C_{\rm P}/T$ vs $T$ data are magnified in the inset of Fig.~\ref{Fig5} to highlight the transition. With the application of external magnetic field, the transition remains unchanged even upto 9~T. In magnetic insulators, $C_{\rm P}$ is the sum of two contributions: lattice/phonon excitations and magnetic contribution.
Typically, $C_{\rm P}$ is dominated by the phonon excitations at high temperatures, while magnetic part dominates at low temperatures if the energy scale of the exchange coupling is low. However, in the compound under investigation, due to the large exchange coupling, the magnetic contribution is extended towards high temperatures. Thus, because of the large exchange coupling and unavailability of an appropriate nonmagnetic analogue compound, it was not possible to subtract the lattice contribution from the total $C_{\rm p}(T)$ to obtain the magnetic heat capacity. Small enhancement of $C_{\rm p}/T$ above $T_{\rm N}$ indicates short-range correlation among the Cu$^{2+}$ ions.

$C_{\rm p}$ well below $T_{\rm N}$ was fitted by a power-law of the form $T^{\alpha}$ in the temperature range 2 to 10~K that yields $\alpha\simeq 2.6$. This value is lower than the expected value of $\alpha=3$ in a 3D AFM ordered state. A reduced value of $\alpha$ has been observed in many frustrated magnets possibly due to the persistence of some fluctuating moments below $T_{\rm N}$~\cite{Zivkovic157204}.

\subsection{$^{51}$V NMR}
In the crystal structure of CBCVO, two adjacent capped-kagome planes are coupled through the V(1)O$_4$ and V(2)O$_4$ units as shown in Fig.~\ref{Fig1}(a). Hence, one can probe the low-lying excitations of Cu$^{2+}$ spins by means of $^{51}$V NMR. 

\subsubsection{$^{51}$V NMR Spectra}
\begin{figure}[h]
 \includegraphics[width=\linewidth]{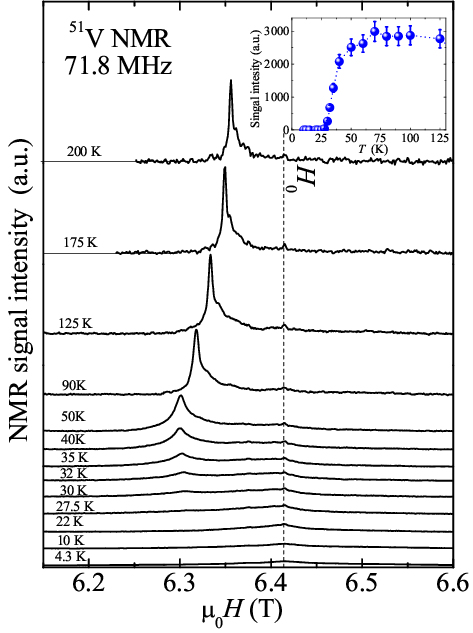}
	\caption {Temperature evolution of the field-swept $^{51}$V NMR spectra measured at a radio frequency of $71.8$~MHz. For clear visualization, the signal intensity for each spectrum was multiplied by temperature, and each spectrum was offset vertically by adding appropriate constants. The vertical dotted line is the zero shift position ($H_{\rm 0}$). The inset shows the temperature dependence of the intrinsic signal intensity which is estimated from the observed intensity multiplied by temperature.}
	\label{Fig6}
\end{figure}
Field-swept $^{51}$V NMR spectra of CBCVO measured in a radio frequency of $71.8$~MHz and different temperatures ($4.3\leq T\leq 250$~K) are presented in Fig.~\ref{Fig6}. The spectrum manifests a single spectral line with an asymmetric shape at high temperatures. As the temperature is lowered, the peak position shifts and the line width increases. Below around 50~K, the signal intensity starts to decrease rapidly as shown in the inset of Fig.~\ref{Fig6}, and then the signal vanishes completely below about 27~K. This is a clear indication of the transition to a magnetic LRO state. In addition to the intrinsic signal, we observed a very broad $^{51}$V NMR signal centered at the zero-shift position. Since the peak position and the signal intensity are almost independent of temperature even above 27~K, we attribute this to an extrinsic signal coming from possible impurities such as V$_2$O$_5$, CuV$_2$O$_5$, etc, though any such phase is not detected in the powder XRD.

$^{51}$V being a quadrupolar nucleus with $I=7/2$, one would expect seven quadruploar-split NMR lines: one central line corresponding to $I_{z} = -1/2 \longleftrightarrow 1/2$ and the six satellite lines corresponding to $I_{z}$ = $\pm 1/2 \longleftrightarrow \pm3/2\longleftrightarrow\pm5/2\longleftrightarrow\pm7/2$ transitions. However, we did not observe any clear satellite peaks, which is probably due to a small value of the quadrupolar frequency ($\nu_{\rm Q}$). In fact, our estimation of $\nu_{\rm Q}$ using a point charge model gives very small values of 0.075~MHz ($\sim 0.0067$~T) and 0.091~MHz ($\sim 0.0081$~T) for V(1) and V(2), respectively. Therefore, we consider that magnetic line broadening easily smears out quadrupolar split lines due to extremely small $\nu_{\rm Q}$ values~\cite{Ranjith024422,Sebastian064413}.
Furthermore, as there are two inequivalent V sites in the crystal structure, the appearance of a single spectral line suggests that both the V sites are almost equivalent or have nearly the same local environment. The asymmetric line can be attributed either to the asymmetry in hyperfine coupling or anisotropy in spin susceptibility or two in-equivalent V sites~\cite{Yogi024413}. It is often possible to access the nature of magnetic ordering by analyzing the NMR spectra on the powder sample below $T_{\rm N}$~\cite{Ranjith014415}. Unfortunately, in our case, it was not possible at all to measure the spectra below $T_{\rm N}$, as the signal vanishes rapidly.

\begin{figure}[h]
\includegraphics[width=\linewidth]{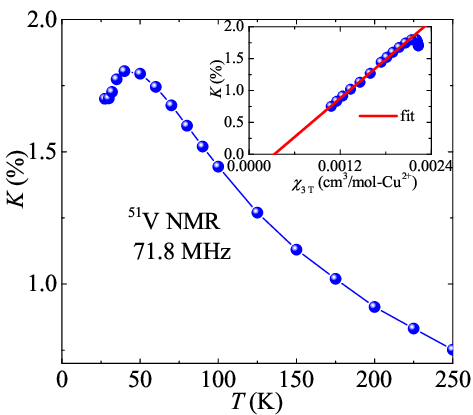}
\caption {Temperature dependent $^{51}$V NMR shift ($K$) measured at 71.8~MHz. Inset: $K$ is plotted against $\chi$ measured at 3~T. The solid line is a straight line fit.}
\label{Fig7}
\end{figure}
Temperature-dependent NMR shift [$K(T)$] of $^{51}$V is extracted from the shift of the central transition ($I_{z} = -1/2 \longleftrightarrow 1/2$) line with respect to $H_{\rm 0}$ and is presented in Fig.~\ref{Fig7}. It passes through a broad maximum at around $T_{\rm K}^{\rm max} \simeq 40$~K, similar to the $\chi(T)$ data, which mimics low-dimensional short-range ordering. $K(T)$ is directly related to the intrinsic local spin susceptibility [$\chi_{\rm spin}(T)$] and is free from extrinsic impurities and defects.
Therefore, $K(T)$ can be expressed in terms of $\chi_{\rm spin}(T)$ by a linear relation
\begin{equation}
	\label{K}
	K(T) = K_0 +\frac{A_{\rm hf }}{N_{\rm A}}\chi_{\rm spin}(T).
\end{equation}
Here, $K_0$ is the temperature-independent chemical (orbital) shift and $A_{\rm hf}$ is the total hyperfine coupling constant between the $^{51}$V nuclear spin and Cu$^{2+}$ electronic spins. $A_{\rm hf}$ is the sum of transferred hyperfine coupling and the dipolar coupling. Usually, dipolar coupling is very small and negligible compared to transferred hyperfine coupling. The $K$ vs $\chi$ is plotted in the inset of Fig.~\ref{Fig7}, where temperature is an indirect variable. We used the $\chi(T)$ data measured at 3~T. It shows a linear behavior down to 50~K and the deviation below 50~K could be due to the presence of the extrinsic contributions in the $\chi(T)$ data. A linear fit to the $K-\chi$ plot in the temperature range 50~K$\leq T\leq 270$~K results in $K_{0} \simeq -0.336\%$ and $A_{\rm hf}\simeq 1.13$~T/$\mu_{\rm B}$. The value of $A_{\rm hf}$ is comparable with the values reported for other vanadates~\cite{Ogloblichev144404}. Such a large value of $A_{\rm hf}$ indicates strong overlap between the $p$ orbital of the V$^{5+}$ ion and $d$ orbital of Cu$^{2+}$ ion through the $p$ orbital of O$^{2-}$. This also implies a significant interaction between the Cu$^{2+}$ ions of adjacent capped-kagome layers through the Cu-O-V-O-Cu pathway.

\subsubsection{Spin-lattice relaxation rate $^{51}1/T_1$}
\begin{figure}[h]
\includegraphics[width=\linewidth]{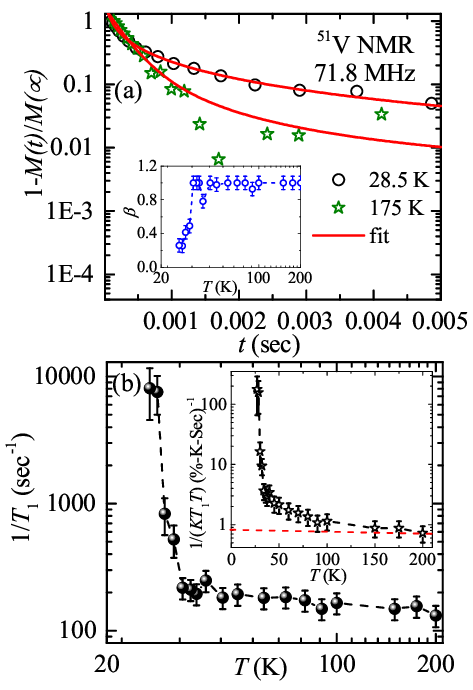}
\caption{(a) Longitudinal magnetization recovery curves at two different temperatures and the red solid lines are the fits using Eq.~\eqref{exp}. Inset: Exponent $\beta$ as a function of $T$. (b) $^{51}$V spin-lattice relaxation rate ($1/T_1$) as a function of $T$. For a clear visualization, the data are shown in log-log scale. Inset: $1/(KT_1 T)$ vs $T$.}
\label{Fig8}
\end{figure}
To understand the spin dynamics, we measured the $^{51}$V spin-lattice relaxation rate ($1/T_1$) as a function of temperature at the central peak position. The recovery of the longitudinal magnetization data shown in Fig.~\ref{Fig8}(a) are fitted by the following stretched multi-exponential function with the stretch exponent $\beta$ ($\leq 1$)~\cite{Gordon783,Simmons1168}: 
\begin{eqnarray}
	1-\frac{M(t)}{M(\infty)} &=& 0.0119\times e^{-(t/T_{1})^{\beta}}+0.068\times e^{-(6t/T_{1})^{\beta}}\nonumber\\
	&+& 0.21\times e^{-(15t/T_{1})^{\beta}}+0.71\times e^{-(28t/T_{1})^{\beta}}.
	\label{exp}
\end{eqnarray}
Here, $M(t)$ is the nuclear magnetization at a time $t$ after the saturation pulse and $M(\infty)$ is the nuclear magnetization at thermal equilibrium. This function is relevant for the central peak of the spectrum of a $I = 7/2$ quadrupole nucleus. For $T > 34$~K, the nuclear recovery curves are well fitted with $\beta$ close to unity. However, below $\sim 34$~K, $\beta$ value decreases down to $\sim 0.4$ indicating the distribution of $T_1$ [see inset of Fig.~\ref{Fig8}(a)].

Temperature dependence of $1/T_1$ extracted from the above fit by Eq.~\eqref{exp} is presented in Fig.~\ref{Fig8}(b). In the high-temperature regime, $1/T_1$ appears to be weakly temperature dependent. A temperature independent behaviour is expected for the uncorrelated spins in the paramagnetic state~\cite{Moriya23}. At low temperatures, $1/T_1$ shows a sharp enhancement below about 32~K associated with the critical slowing down of the fluctuating moments as one approaches the magnetic ordering temperature ($T_{\rm N}$). Unfortunately, due to the disappearance of the NMR signal from the measurement window, we couldn't measure $1/T_1$ below 27~K. This suggests that the transition temperature at 6.3~T ($\sim 71.8$~MHz) is lower than 27~K. Indeed, no shift in $T_{\rm N}$ is observed from the heat capacity measurements with field upto 9~T~\cite{Nath214430,Nath064422}.
It is further noted that the enhancement of $1/T_1$ below 32~K is not due to the extrinsic signal as the $T_1$ of the extrinsic signal is more than a few order of magnitude longer than the measured short $T_1$.
  

$\frac{1}{T_{1}}$ in terms of the dynamic susceptibility $\chi_{M}(\vec{q},\omega_{0})$ can be written as~\cite{Moriya516}
\begin{equation}
	\frac{1}{T_{1}T} = \frac{2\gamma_{N}^{2}k_{B}}{N_{\rm A}^{2}}
	\sum\limits_{\vec{q}}\mid A(\vec{q})\mid
	^{2}\frac{\chi^{''}_{M}(\vec{q},\omega_{0})}{\omega_0},
 	\label{t1form}
 \end{equation}
 where the sum is over the wave vector $\vec{q}$ within the first Brillouin zone, $A(\vec{q})$ is the $q$-dependent form-factor of the hyperfine interaction, and $\chi^{''}_{M}(\vec{q},\omega_{0})$ is the imaginary part of the dynamic susceptibility at the wave vector $\vec{q}$ and the nuclear Larmor frequency $\omega _0$. For $q=0$ and $\omega_{0}=0$, the real component of $\chi_{M}^{'}(\vec{q},\omega _{0})$ corresponds to the uniform static susceptibility $\chi$. For a dominant contribution of $\chi$ to $1/T_{1}T$, $1/(\chi T_{1}T)$ should show temperature independent behaviour. As depicted in the inset of Fig.~\ref{Fig8}(b), $1/(KT_{1}T)$ is not exactly constant at high temperatures, rather, it exhibits a slight increase with the reduction in temperature. This demonstrates the growth of antiferromagnetic correlations with decreasing temperature or persistence of magnetic correlations upto high temperatures, which is consistent with the large value of $\theta_{\rm CW}$.

In the high-temperature limit ($T>>\theta_{\rm CW}$), by considering the Gaussian form of the auto-correlation function, one can get~\cite{Moriya23,Ranjith024422} 
\begin{eqnarray}
	\begin{split}
		\lefteqn{\left(\frac{1}{T_{1}}\right)_{T\to \infty} = \frac{(\gamma_{N}g\mu_{\rm B})^{2}\sqrt{2\pi}z^\prime S(S+1)}{3\omega_{\rm ex}}}\\&\qquad&\qquad&\qquad
		\times\left(\frac{A_{hf}}{z^\prime}\right)^{2},
	\end{split}
	\label{t1form_2}
\end{eqnarray}
where $\omega_{\rm ex}=(|J^{\rm max}|k_{\rm B}/\hbar)\sqrt{\frac{2zS(S+1)}{3}}$ is the Heisenberg exchange frequency, $z$ is the number of NN spins of each Cu$^{2+}$ ion, and $z^\prime$ is the number of NN Cu$^{2+}$ spins of the V site. In the above expression, total hyperfine coupling is divided by $z^\prime$ in order to account for the coupling of V site with the individual Cu$^{2+}$ ion.
In the crystal structure, each V site is strongly coupled with six NN Cu$^{2+}$ ions. Using the appropriate parameters [$A_{\rm hf} \simeq 1.13$~T/$\mu_{\rm B}$, $\gamma_N = 70.76\times10^2$~rad.sec$^{-1}$/Oe, $z=4.5$, $z^{\prime}=6$, $g \simeq 2.29$ [obtained from the $\chi(T)$ analysis], $S= 1/2$, and the relaxation rate at 200~K $\left(\frac{1}{T_{1}}\right)_{T\to \infty}\simeq 131$ sec$^{-1}$], the magnitude of the leading AFM exchange coupling between Cu$^{2+}$ ions is estimated to be $J^{\max}/k_{\rm B}\simeq136$~K which agrees with the value obtained from the $\theta_{\rm CW}$ analysis. This value is indeed comparable with the leading exchange coupling reported for CCCVO~\cite{Botana054421}.

\section{Summary} 
The value of the frustration parameter of CBCVO, $f \simeq 8$ indicates that the system is strongly frustrated. This value is comparable to that reported for CCCVO ($f \simeq 7.7$)~\cite{Winiarski58,Botana054421}. As the exchange coupling between the Cu$^{2+}$ ions is strongly dependent on the Cu-Cu bond distances and $\angle$Cu-O-Cu bond angles, in the following, we made a comparison of the structural aspects of these two compounds. 
As mentioned earlier, there is an inconsistency in the crystal structure of CCCVO extracted from powder and single crystal samples. The synchrotron experiments on the powder sample reveal a structural phase transition from the high-temperature trigonal ($P\bar{3}m1$) to low-temperature monoclinic ($P2_1/c$) structure at $\sim 310$~K~\cite{Botana054421}.
On the other hand, the structural data reported on the single crystal suggest a trigonal structure with a different space group ($P\bar{3}$) even at room temperature ($\sim 296$~K)~\cite{Kornyakov1833}. We have adopted the single crystal data of CCCVO as a reference to refine the powder XRD data of CBCVO.
In both CCCVO and CBCVO, CuO$_4$ plaquette and CuO$_5$ polyhedra are distorted.
In CCCVO, the Cu-Cu bond lengths and $\angle$Cu-O-Cu within the triangular base of OCu$_4$ unit are in the range of 3.18-3.19~\AA~and 116-117$^{\degree}$, respectively. On the other hand for CBCVO, these values lie in between 2.95-3.41~\AA~and 119.5-122.5$^{\degree}$, respectively.
Similarly, in CCCVO, the Cu-Cu bond lengths and $\angle$Cu-O-Cu between the Cu$^{2+}$ ions in the kagome layer and in the capped position are in the rage of 2.92-3.18~\AA~and 101-102$^{\degree}$, respectively. For CBCVO, theses values vary in the range of 2.8-3.06~\AA~and 98-101$^{\degree}$, respectively. The above assessment reflects nearly equal bond distances and bond angles implying that the exchange interactions among the Cu$^{2+}$ ions in both CCCVO and CBCVO compounds should be comparable. Interestingly, no signature of structural transition from high temperature trigonal to low temperature monoclinic (lower symmetry) structure is detected for CBCVO which is in sharp contrast to that reported for CCCVO on the powder sample. Further, the appearance of an extra peak below 130~K in the powder XRD data bear a resemblance to CCCVO, the orgin of which is not known yet and requires further investigations.

In summary, we have synthesized and studied the magnetic properties of a capped-kagome compound CBCVO by performing $\chi(T)$, $C_{\rm p}(T)$, and $^{51}$V NMR measurements. The magnetic LRO sets in at $T_{\rm N} \simeq 21.5$~K in zero magnetic field, which is well below the CW temperature $\theta_{\rm CW} \simeq -175$~K. This is a footprint of strong magnetic frustration in the compound under investigation. A clear signature of the onset of magnetic LRO is confirmed from the sudden disappearance of NMR signal and divergence of $1/T_1(T)$.
The strong hyperfine coupling of $^{51}$V nucleus with the magnetic Cu$^{2+}$ ions implies significant overlapping of orbitals that facilitates strong AFM exchange coupling via Cu-O-V-O-Cu route. Further experiments on a good quality single crystal would reveal the exact nature of the ground state.

\acknowledgments
We would like to acknowledge SERB, India for financial support bearing sanction Grant No. CRG/2022/000997. SG was supported by the Prime Minister’s Research Fellowship (PMRF) scheme, Government of India. Work at the Ames National Laboratory was supported by the U.S. Department of Energy, Office of Science, Basic Energy Sciences, Materials Sciences and Engineering Division. The Ames Laboratory is operated for the U.S. Department of Energy by Iowa State University under Contract No.~DEAC02-07CH11358.

%

\end{document}